\documentstyle[aps,pre,preprint,tighten]{revtex}  % for tighten preprint

\begin{document}
\draft

\title{Random Matrix Theory and Classical Statistical Mechanics.
I. Vertex Models}

\author{H.\ Meyer\cite{byline}, J.-C.\ Angl\`es d'Auriac\cite{byline}}
\address{
Centre de Recherches sur les Tr\`es Basses Temp\'eratures,\\
BP 166, 38042 Grenoble, France}

\author{J.-M.\ Maillard\cite{bymaillard}}
\address{Laboratoire de Physique Th\'eorique et Hautes Energies \\
Tour 16, 1$^{\rm er}$ \'etage, 4 place Jussieu, 
75252 Paris Cedex, France}

\date{\today}

\maketitle
\hfill {PAR--LPTHE 96--35}

\begin{abstract}
A connection between integrability properties and general
statistical properties of the spectra of symmetric transfer matrices
of the asymmetric eight-vertex model is studied using random
matrix theory (eigenvalue spacing distribution and spectral
rigidity). For Yang-Baxter integrable cases,
including free-fermion solutions, we have found a Poissonian
behavior, whereas level repulsion close to the Wigner
distribution is found for non-integrable models. 
For the asymmetric eight-vertex model, however,
the level repulsion can also disappear
and the Poisson distribution be recovered
on (non Yang--Baxter integrable) algebraic varieties,
the so-called disorder varieties.
We also present an infinite set of algebraic
varieties which are stable under the action of
an infinite discrete symmetry group of the parameter space.
These varieties are possible loci for free parafermions.
Using our numerical criterion we have tested the generic
calculability of the model on these algebraic varieties.
\end{abstract}


\section{Introduction}
Since the early work of Wigner \cite{Wig53}
random matrix theory (RMT) has been applied with success in many
domains of physics~\cite{Meh91}.
Initially developed to serve for nuclear physics, RMT proves itself
to provide an adequate description to any situation implying chaos.
It has been found that the spectra of many quantum systems is very close
to one of four archetypal situations described by four statistical
ensembles. For the few integrable models this is the ensemble of
diagonal random matrices, while for non-integrable systems this can be
the Gaussian Orthogonal Ensemble (GOE),
the Gaussian Unitary Ensemble (GUE), or
the Gaussian Symplectic Ensemble (GSE),
depending on the symmetries of the model under consideration.

In the last years several quantum spin Hamiltonians have been investigated
from this point of view.
It has been found \cite{PoZiBeMiMo93,HsAdA93}
that 1D systems for which the Bethe
ansatz applies have a level spacing distribution
close to a Poissonian (exponential) distribution, $P(s) = \exp(-s)$,
whereas if the Bethe ansatz does not apply, the level spacing distribution
is described by the Wigner surmise for the Gaussian orthogonal ensemble (GOE):
\begin{equation}
\label{e:wigner}
P(s) = \frac{\pi}{2} s \exp( -\pi s^2 / 4) \;.
\end{equation}
Similar results have been found for 2D quantum spin systems
\cite{MoPoBeSi93,vEGa94,BrAdA96}.
Other statistical properties
have also been analyzed, showing that the description of 
the spectrum of the quantum spin system
by a statistical ensemble is valid not only 
for the level spacings but also
for quantities involving more than two eigenvalues.

In a recent letter \cite{hm4} we proposed the 
extension of random matrix theory
analysis to models of classical statistical mechanics
(vertex and spin models), studying
the transfer matrix of the eight-vertex model as an example.
The underlying idea is that, if there actually exists
a close relation between integrability and the Poissonian
character of the distribution, it could be
better understood in a framework which makes
Yang--Baxter integrability and its key structures 
(commutation of transfer matrices depending on 
spectral parameters) crystal clear: one wants to
switch from quantum Hamiltonian framework to
transfer matrix framework.
We now present the complete results of our study of transfer matrices
and a detailed description of the numerical method.
This work is split into two papers:
the first one describes the
numerical methods and the results on the eight-vertex model,
the second one treats the case of discrete spin models with the
example of the Ising model in two and three dimensions and
the standard Potts model with three states.

We will analyze  a possible connection between
statistical properties of the entire spectrum of the model's
transfer matrix and the Yang--Baxter integrability.
A priori, such a connection is not sure to exist since only the few
eigenvalues with largest modulus have a physical signification,
while we are looking for properties of the entire spectrum.
However, our numerical results 
show a connection which we will discuss.
We will also give an extension of the so-called ``disorder 
variety'' to the asymmetric eight-vertex model 
where the partition
function can be summed up without Yang--Baxter integrability.
We then present an infinite discrete symmetry group
of the model and an infinite set of
algebraic varieties stable under this group.
Finally, we test all these varieties
from the point of view of RMT analysis.

This paper is organized as follows: 
in Sec.~\ref{s:numeric} we recall the
machinery of RMT, and we give some details 
about the numerical methods we use.
Sec.~\ref{s:8v} is devoted to the eight-vertex model. 
We list the cases where the partition function can be
summed up, and give some new analytical results concerning
the disorder variety and the automorphy group
of the asymmetric eight-vertex model.
The numerical results of the analysis of the spectrum 
of transfer matrices 
are presented in Sec.~\ref{s:results8v}.
The last section concludes with a discussion.

\section{Numerical Methods of RMT}
\label{s:numeric}

\subsection{Unfolding of the Spectrum}

\begin{figure}
\psfig{file=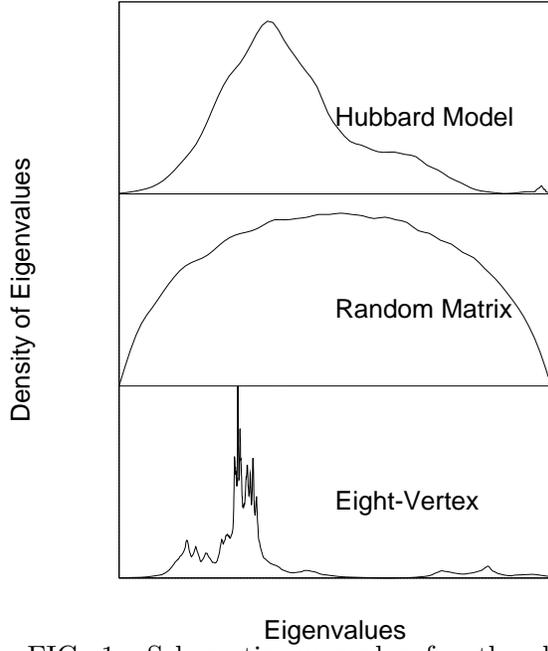,width=\hsize}
\caption{Schematic examples for the density of eigenvalues 
for the 2d Hubbard, GOE random matrices, 
and transfer matrices of the eight-vertex model.}
\label{f:density}
\end{figure}

In RMT analysis
one considers the spectrum of the (quantum) Hamiltonian, or of the
transfer matrix, as a
collection of numbers, and one looks for some possibly universal
statistical properties of this collection of numbers.
Obviously, the raw spectrum will not have any universal properties.
For example, Fig.~\ref{f:density} shows schematically
three densities of eigenvalues:
for a 2d Hubbard model, for an eight-vertex model and for the
Gaussian Orthogonal Ensemble. They have clearly nothing in common.
To find universal properties, one has to perform a kind of renormalization
of the spectrum, this is the so-called unfolding operation.
This amounts to making the {\em local} density of eigenvalues equal to
unity everywhere in the spectrum.
In other words, one has to subtract the regular part from the integrated
density of states and consider only the fluctuations.
This can be achieved by different means, however, there is no rigorous
prescription and the best criterion is the insensitivity of the final
result to the method employed or to the parameters 
(for ``reasonable'' variation).

Throughout this paper, we call $E_i$ the raw eigenvalues and
$\epsilon_i$ the corresponding unfolded eigenvalues.
Thus the requirement is that the local density of the $\epsilon_i$'s is one.
We need to compute an averaged integrated density of states $\bar\rho(E)$
from the actual integrated density of states:
\begin{equation}
\rho(E)={1\over N}\int_{-\infty}^E \sum_i{\delta(e-E_i)}\,de \;,
\end{equation}
and then we take $\epsilon_i = N \bar\rho(E_i)$.
To compute $\bar\rho(E)$ from $\rho(E)$, we have performed a running average:
we choose some odd integer $2r+1$ of the order of 9--25 and then replace
each eigenvalue $E_i$ by a local average:
\begin{equation}
E_i^\prime = {1\over 2r+1} \sum_{j=i-r}^{i+r} E_j  \;,
\end{equation}
and $\bar\rho(E)$ is approximated by the linear interpolation between
the points of coordinates $(E_i^\prime,i)$.
We compared the results with other methods:
one can replace
each delta peak in $\rho(E)$ by a Gaussian with a properly chosen mean square
deviation. Another method is to discard the low frequency components in
a Fourier transform of $\rho(E)$.
A detailed explanation and tests of these methods of unfolding are given
in Ref.~\cite{BrAdA97}.
Note also that for very peculiar spectra it is necessary to break it into
parts and to unfold each part separately. Also the extremal eigenvalues
are discarded since they induce finite size effects.
It comes out that of the
three methods, the running average unfolding is the best suited in
the context of transfer matrices, and it is also the fastest.

\subsection{Symmetries}

For quantum Hamiltonians, it is well known that
it is necessary to sort the eigenvalues with
respect to their quantum numbers, and to compare only
eigenvalues of states belonging to the same quantum numbers.
This is due to the fact that eigenstates with different symmetries are
essentially uncorrelated.
The same holds for transfer matrices.
In general, a transfer matrix $T$ of a classical 
statistical mechanics lattice model (vertex model)
depends on several parameters (Boltzmann weights $w_i$). Due to
the lattice symmetries, or to other symmetries (permutation of colors
and so on), there exist some operators $S$ acting on the same space as the 
transfer matrix and which are {\em independent of the parameters},
commuting with $T$: $[T(\{w_i\}),S] = 0$.
It is then possible to find subspaces of $T$ which are also
independent of the parameters.
Projection on these invariant subspaces amounts to block-diagonalizing $T$
and to split the unique spectrum of $T$ into the many spectra of each block.
The construction of the projectors is done with the help of the character
table of irreducible representations of the symmetry group.
Details can be found in \cite{BrAdA97,hmth}.

As we will discuss in the next sections, we always restricted ourselves
to symmetric transfer matrices.
Consequently the blocks are also symmetric and there are only {\em real}
eigenvalues. The diagonalization is performed
using standard methods of linear algebra (contained in the LAPACK library).
The construction of the transfer matrix and
the determination of its symmetries
depend on the model and are detailed in Sec.~\ref{s:transfer}
for the eight-vertex model.

\subsection{Quantities Characterizing the Spectrum}
\label{s:quantities}

Once the spectrum has been obtained and unfolded, various statistical
properties of the spectrum are investigated. The simplest one is the
distribution $P(s)$ of the spacings $s=\epsilon_{i+1}-\epsilon_i$
between two consecutive unfolded eigenvalues.
This distribution will be compared to an exponential 
and to the Wigner law (\ref{e:wigner}).
Usually, a simple visual inspection is sufficient to recognize
the presence of level repulsion, the main property for non-integrable
models.
However, to quantify the ``degree'' of level repulsion, it is convenient
to use  a parameterized distribution which interpolates between the
Poisson law, the Wigner law.
From the many possible distributions we have chosen
the Brody distribution \cite[ch.\ 16.8]{Meh91}:
\begin{mathletters}
\begin{equation}
P_\beta(s) = c_1\, s^\beta\, \exp\left(-c_2 s^{\beta+1}\right)
\end{equation}
with 
\begin{equation}
c_2=\left[\Gamma\left({\beta+2\over\beta+1}\right)\right]^{1+\beta}
\quad\mbox{and}\quad  c_1=(1+\beta)c_2 \;.
%% metha chap 16.8
\end{equation}
\end{mathletters}
For $\beta=0$, this is a simple exponential for the Poisson ensemble,
and for $\beta=1$, one recovers the Wigner surmise for the GOE.
This distribution turns out to be convenient since its indefinite
integral can be expressed with elementary functions.
It has been widely used in the literature, except when special
distributions were expected as at the metal insulator transition
\cite{VaHoScPi95}.
Minimizing the quantity:
\begin{equation}
\phi(\beta) = \int_0^\infty(P_\beta(s)-P(s))^2 \,ds
\end{equation}
yields a value of $\beta$ characterizing the degree of level repulsion
of the distribution $P(s)$. We have always found $\phi(\beta)$ small.
When $-0.1<\beta<0.1$, the distribution is close to a Poisson law,
while for $0.5<\beta<1.2$ the distribution is close to the Wigner surmise.

If a distribution is found to be close to the Wigner surmise (or the
Poisson law), this does not mean that the GOE (or the Diagonal Matrices
Ensemble) describes correctly the spectrum.
Therefore it is of interest to compute functions involving higher
order correlations as for example the spectral rigidity
\cite{Meh91}:
\begin{equation}
\Delta_3(E) = 
\left\langle \frac{1}{E} \min_{a,b}
\int_{\alpha-E/2}^{\alpha+E/2}
{\left( N(\epsilon)-a \epsilon -b\right)^2 d\epsilon} \right\rangle_\alpha \;,
\end{equation}
where $\langle\dots\rangle_\alpha$ denotes an average over 
the whole spectrum.
This quantity measures the deviation from equal spacing.
For a totally rigid spectrum, as that of the harmonic oscillator, one has
$\Delta_3^{\rm osc}(E) = 1/12$, for an integrable (Poissonian) system one has
$\Delta_3^{\rm Poi}(E) = E/15$, while for the Gaussian Orthogonal
Ensemble one has
$\Delta_3^{\rm GOE}(E) = \frac{1}{\pi^2} (\log(E) - 0.0687) + {\cal O}(E^{-1})$.
It has been found that the spectral rigidity of  quantum spin systems 
follows $\Delta_3^{\rm Poi}(E)$ in the integrable case and
$\Delta_3^{\rm{GOE}}(E)$ in the non-integrable case.
However, in both cases, even though $P(s)$ is in good agreement with RMT,
 deviations from RMT occur for $\Delta_3(E)$ at some system dependent
point $E^*$.
This stems from the fact that the rigidity
$\Delta_3(E)$ probes correlations beyond nearest neighbours
 in contrast to $P(s)$.
%This is probably why the rigidity is much more sensitive
%to the parameters of the unfolding than the spacing distribution
%as mentioned in section \ref{s:results8v}.

\section{The Asymmetric Eight-Vertex Model on a square lattice}
\label{s:8v}
\subsection{Generalities}

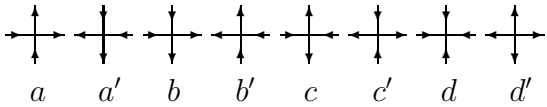
\begin{figure}

\centerline{
%\vertex{-}{-}{-}{-}{a}%
\begin{picture}(22,40)(-11,-30)
\put(-11,0){\line(1,0){22}}
\put(0,-11){\line(0,1){22}}
\put(-8,0){\vector(--1,0){3}}
\put(8,0){\vector(--1,0){3}}
\put(0,-8){\vector(0,--1){3}}
\put(0,8){\vector(0,--1){3}}
\put(-2,-25){$a$}
\end{picture}
%\vertex{-}{+}{+}{-}{a'}%
\begin{picture}(22,40)(-11,-30)
\put(-11,0){\line(1,0){22}}
\put(0,-11){\line(0,1){22}}
\put(-8,0){\vector(-+1,0){3}}
\put(8,0){\vector(+-1,0){3}}
\put(0,-8){\vector(0,+-1){3}}
\put(0,8){\vector(0,-+1){3}}
\put(-2,-25){$a'$}
\end{picture}
%\vertex{+}{-}{+}{-}{b}%
\begin{picture}(22,40)(-11,-30)
\put(-11,0){\line(1,0){22}}
\put(0,-11){\line(0,1){22}}
\put(-8,0){\vector(++1,0){3}}
\put(8,0){\vector(--1,0){3}}
\put(0,-8){\vector(0,+-1){3}}
\put(0,8){\vector(0,+-1){3}}
\put(-2,-25){$b$}
\end{picture}
%\vertex{-}{-}{+}{+}{b'}%
\begin{picture}(22,40)(-11,-30)
\put(-11,0){\line(1,0){22}}
\put(0,-11){\line(0,1){22}}
\put(-8,0){\vector(-+1,0){3}}
\put(8,0){\vector(-+1,0){3}}
\put(0,-8){\vector(0,++1){3}}
\put(0,8){\vector(0,--1){3}}
\put(-2,-25){$b'$}
\end{picture}
%\vertex{-}{-}{-}{+}{c}%
\begin{picture}(22,40)(-11,-30)
\put(-11,0){\line(1,0){22}}
\put(0,-11){\line(0,1){22}}
\put(-8,0){\vector(--1,0){3}}
\put(8,0){\vector(-+1,0){3}}
\put(0,-8){\vector(0,-+1){3}}
\put(0,8){\vector(0,--1){3}}
\put(-2,-25){$c$}
\end{picture}
%\vertex{+}{-}{-}{-}{c'}%
\begin{picture}(22,40)(-11,-30)
\put(-11,0){\line(1,0){22}}
\put(0,-11){\line(0,1){22}}
\put(-8,0){\vector(+-1,0){3}}
\put(8,0){\vector(--1,0){3}}
\put(0,-8){\vector(0,--1){3}}
\put(0,8){\vector(0,+-1){3}}
\put(-2,-25){$c'$}
\end{picture}
%\vertex{-}{+}{-}{-}{d}%
\begin{picture}(22,40)(-11,-30)
\put(-11,0){\line(1,0){22}}
\put(0,-11){\line(0,1){22}}
\put(-8,0){\vector(--1,0){3}}
\put(8,0){\vector(+-1,0){3}}
\put(0,-8){\vector(0,--1){3}}
\put(0,8){\vector(0,-+1){3}}
\put(-2,-25){$d$}
\end{picture}
%\vertex{-}{-}{+}{-}{d'}}
\begin{picture}(22,40)(-11,-30)
\put(-11,0){\line(1,0){22}}
\put(0,-11){\line(0,1){22}}
\put(-8,0){\vector(-+1,0){3}}
\put(8,0){\vector(--1,0){3}}
\put(0,-8){\vector(0,+-1){3}}
\put(0,8){\vector(0,--1){3}}
\put(-2,-25){$d'$}
\end{picture}}

\caption{The Boltzmann weights of the eight vertices.}
\label{f:vertices}
\end{figure}

We will focus in this section on the asymmetric 
eight-vertex model on a square lattice.
We use the standard notations of Ref.~\cite{Bax82}.
The eight-vertex condition specifies that only vertices are allowed
which have an even number of arrows pointing to the center of the
vertex.
Fig.~\ref{f:vertices}
shows the eight vertices with their corresponding Boltzmann weight.
The partition function per site depends on these eight homogeneous variables 
(or equivalently seven independent values):
\begin{equation}
Z(a,a',b,b',c,c',d,d')\;.
\end{equation}
It is customary to arrange the eight (homogeneous) Boltzmann weights
in a $4 \times 4$ $R$-matrix:
\begin{eqnarray}
\label{e:Rmat}
{\cal R} \, = \,
 \left(
\begin {array}{cccc} 
a&0&0&d\\
0&b& c&0\\
0&c^\prime&b^\prime&0\\
d^\prime&0&0& a^\prime
\end {array}
\right)
\end{eqnarray}
The entry ${\cal R}_{i j}$ is the Boltzmann weight
of the vertex defined by the four digits of the binary representation
of the two indices $i$ and $j$. The row index corresponds
to the east and south edges and the column index corresponds
to the west and north edges:
\[
{\cal R}_{i j}={\cal R}_{\mu\alpha}^{\nu\beta}
=w(\mu,\alpha|\beta,\nu)
\]
\[
\begin{picture}(40,30)(-20,-15)
\put(-10,0){\line(1,0){20}}
\put(0,-10){\line(0,1){20}}
%\put(-15,0){\vector(1,0){3}}
%\put(15,0){\vector(1,0){3}}
%\put(0,-15){\vector(0,1){3}}
%\put(0,15){\vector(0,1){3}}
\put(-20,-3){$\mu$}
\put(-3,13){$\beta$}
\put(13,-3){$\nu$}
\put(-3,-20){$\alpha$}
\end{picture}
\]
When the Boltzmann weights are 
unchanged by negating all the four edge values the model
is said {\em symmetric} otherwise it is {\em asymmetric}. This should 
not be confused with the symmetry of the transfer matrix.
Let us now discuss a general symmetry property of the model.
A combinatorial argument \cite{Bax82} shows that for any lattice
without dangling ends,
the two parameters $c$ and $c^\prime$ can be taken equal, and
that, for most regular lattices (including the periodic
square lattice considered
in this work), $d$ and $d^\prime$ can 
also be taken equal (gauge transformation \cite{GaHi75}).
Specifically, one has:
\begin{equation}\label{e:gauge}
Z(a,a',b,b',c,c',d,d') =
Z(a,a',b,b',\sqrt{cc'},\sqrt{cc'},\sqrt{dd'},\sqrt{dd'}) \;.
\end{equation}
We will therefore always take $c=c'$ and $d=d'$ in the
numerical calculations.
In the following, when $c'$ and $d'$ are not mentioned it is
implicitly meant that $c'=c$ and $d'=d$.
Let us finally recall that the 
asymmetric eight-vertex model is equivalent to an Ising
spin model on a square lattice including next nearest neighbor interactions
on the diagonals and four-spin interactions around a plaquette
(IRF model) \cite{Bax82,Kas75}.
However, this equivalence is not exact on a finite lattice since
the $L\times M$ plaquettes do not form a basis
(to have a cycle basis, one must take any $L\times M -1$ plaquettes plus
one horizontal and one vertical cycle).

\subsection{The Row-To-Row Transfer Matrix}
\label{s:transfer}

Our aim is to study the full spectrum of the transfer matrix.
More specifically,
we investigate the properties of the row-to-row transfer
matrix which corresponds to build up a periodic
$L\times M$ rectangular lattice by adding rows of length $L$.
The transfer matrix $T_L$ is a
$2^L\times 2^L$ matrix and the partition function becomes:
\begin{equation}
Z(a,a',b,b',c,d) = {\rm Tr} \, [T_L(a,a',b,b',c,d)]^M\;.
\label{e:Ztrace}
\end{equation}
However, there are many other possibilities to build up the lattice,
each corresponding to another form of transfer matrix: it just has to
lead to the same partition function.
Other widely used examples are 
diagonal(-to-diagonal) and corner transfer matrices \cite{Bax82}.

The index of the row-to-row transfer matrix enumerates the $2^L$
possible configurations of one row of $L$ vertical bonds. We choose a
binary coding:
\begin{equation}
\alpha=\sum_{i=0}^{L-1} \alpha_i2^i \equiv | \alpha_0,\dots,\alpha_{L-1} \rangle
\end{equation}
with $\alpha_i\in\{0,1\}$, 0 corresponding to arrows pointing up or to the right
and 1 for the other directions.
One entry $T_{\alpha,\beta}$ thus describes the contribution to
the partition function of two neighboring rows having the configurations
$\alpha$ and $\beta$:
\begin{equation}
T_{\alpha,\beta} = \sum_{\{\mu\}}
  \prod_{i=0}^{L-1} w(\mu_i,\alpha_i | \beta_i,\mu_{i+1}) \;. \label{e:Tab}
\end{equation}
%
%$\mu_i\alpha_i\beta_i\mu_{i+1}=+1$ IMCOMPATIBLE AVEC $a_i=0,1$ !!
With our binary notation, the eight-vertex condition means that
$w(\mu_i,\alpha_i | \beta_i,\mu_{i+1})=0$ if the sum
$\mu_i+\alpha_i+\beta_i+\mu_{i+1}$ is odd. 
Therefore, the sum (\ref{e:Tab}) reduces to
exactly two terms: once $\mu_0$ is chosen (two possibilities),
$\mu_1$ is uniquely defined since $\alpha_0$ and $\beta_0$ are fixed
and so on.
For periodic boundary conditions,
the entry $T_{\alpha,\beta}$ is zero
if the sum of all $\beta_i$ and $\alpha_i$ is odd.
This property naturally splits the transfer matrix into two blocks:
entries between row configurations with an even number of up arrows
and entries between configurations with an odd number of up arrows.

\subsubsection{Symmetries of the Transfer Matrix}
\label{s:c=d}
Let us now discuss various symmetry properties of the transfer matrix.

(i) When one exchanges the rows $\alpha$ and $\beta$, the vertices of type
$a$, $a'$, $b$, and $b'$ will remain unchanged while the vertices
of type $c$ and $d$ will exchange into one another.
Thus for $c=d$ the transfer matrix $T_L(a,a',b,b',c,d)$ is symmetric.
In general the symmetry of the row-to-row transfer matrix
is satisfied for $c=d'$ and $d=c'$.
In terms of the equivalent IRF Ising model, condition $c=d$ means
that the two diagonal interactions $J$ 
and $J'$ (confer to Ref.~\cite{Bax82})
are the same: the Ising model is isotropic and therefore its 
row-to-row transfer matrix is symmetric, too.
This coincidence is remarkable since the equivalence between
the asymmetric
eight-vertex model and the Ising model is not exact on a finite
lattice as already mentioned.

(ii) We now consider the effect of permutations of lattice sites preserving
the neighboring relations.
Denote by $S$ a translation operator defined by:
\begin{equation}
S|\alpha_0,\alpha_1,\dots,\alpha_{L-1} \rangle
= |\alpha_1,\dots,\alpha_{L-1},\alpha_0\rangle \;.
\end{equation}
Then we have:
\begin{equation}
\langle\alpha S^{-1}|T_L(a,a',b,b',c,d)|S\beta\rangle = 
\langle\alpha|T_L(a,a',b,b',c,d)|\beta\rangle \;,
\end{equation}
and therefore:
\begin{equation}
[T_L(a,a',b,b',c,d),S] = 0 \;.
\end{equation}
For the reflection operator $R$ defined by:
\begin{equation}
R|\alpha_0,\alpha_1,\dots,\alpha_{L-1} \rangle
= |\alpha_{L-1},\dots,\alpha_{1},\alpha_0\rangle \;,
\end{equation}
we have:
\begin{equation}
\langle\alpha R^{-1}|T_L(a,a',b,b',c,d)|R\beta\rangle = 
\langle\alpha|T_L(a,a',b,b',d,c)|\beta\rangle \;.
\end{equation}
Thus $R$ commutes with $T$ only for the symmetric case $c=d$:
\begin{equation}
[T_L(a,a',b,b',c,c),R] = 0 \;.
\end{equation}
Combination of the translations $S$ and the reflection $R$ leads to 
the dihedral group ${\cal D}_L$.
These are all the general lattice symmetries in the square
lattice case.
The one dimensional nature of the group ${\cal D}_L$
reflects the dimensionality of the rows added to the
lattice by a multiplication by $T$. This is general :
the symmetries of the transfer matrices of $d$-dimensional lattice 
models are the symmetries of ($d-1$)-dimensional space.
The translational invariance in the last space direction has already
been exploited with the use of the transfer matrix itself leading to
Eq.~(\ref{e:Ztrace}).

(iii) Lastly, we look at symmetries due to operations on the dynamic
variables themselves.
There is a priori no continuous symmetry in this model 
in contrast with the
Heisenberg quantum chain which has a continuous $SU(2)$ spin symmetry.
But one can define an operator $C$ returning all arrows:
\begin{equation}
C|\alpha_0,\alpha_1,\dots,\alpha_{L-1} \rangle
= |1-\alpha_0, 1-\alpha_1,\dots,1-\alpha_{L-1}\rangle \;.
\end{equation}
This leads to an exchange of primed and unprimed Boltzmann weights:
\begin{equation}
\langle\alpha C^{-1}|T_L(a,a',b,b',c,d)|C\beta\rangle = 
\langle\alpha|T_L(a',a,b',b,c,d)|\beta\rangle \;,
\end{equation}
Thus for the symmetric eight-vertex model (Baxter model)
the symmetry operator $C$ commutes with the transfer matrix:
\begin{equation}
[T_L(a,a,b,b,c,d),C] = 0 \;.
\end{equation}

\subsubsection{Projectors}

Once the symmetries have been identified,
it is simple to construct the projectors of one row of each irreducible
representation of the group ${\cal D}_L$
(details can be found in \cite{BrAdA97,hmth}).
When $L$ is even, there are four representations of dimension 1 and 
$L/2-1$ representations of dimension 2 (i.e.\ in all there are $L/2+3$
projectors). When $L$ is odd, there are two one-dimensional representations
and $(L-1)/2$ representations of dimension 2, in all $(L-1)/2 + 2$ projectors.

For the symmetric model with $a=a'$ and $b=b'$,
there is an extra ${\cal Z}_2$ symmetry
which doubles the number of projectors.

Using the projectors block diagonalizes the transfer matrix leaving a
collection of small matrices to diagonalize instead of the large one.
For example, for $L=16$, the total row-to-row
transfer matrix has the linear size
$2^L=65536$, the projected blocks have linear sizes between 906 and 2065
(see also Tabs.~\ref{t:aj14} and \ref{t:aj16}).

As already mentioned, the block projection not only saves computing time
for the diagonalization but is necessary to sort  the eigenvalues
with respect to the symmetry of the corresponding eigenstates.
In summary, when $c=d$, the row-to-row
transfer matrix is symmetric leading to
a real spectrum. Its symmetries have been identified.
This is a fortunate situation since restriction $c=d$ does neither
prevent, nor enforce, Yang--Baxter integrability as will be 
explained in the following section.

\subsection{Integrability of the Eight-Vertex Model}

We now summarize the cases where the partition function of the
eight-vertex model can be analyzed and possibly computed.
These are the symmetric eight-vertex model,
the asymmetric six-vertex model, the free-fermion variety and some 
``disorder solutions''.

\subsubsection{The Symmetric Eight-Vertex Model}
\label{ss:s8v}

Firstly, in the absence of an `electrical field', i.e.\ when
$a=a'$, $b=b'$, $c=c'$, and $d=d'$, the transfer matrix can be
diagonalized using the Bethe ansatz
or the Yang--Baxter equations \cite{Bax82}.
This case is called the symmetric eight-vertex model, also called
Baxter model \cite{Bax82}.
One finds that two row-to-row transfer matrices $T_L(a,b,c,d)$ and
$T_L(\bar a,\bar b,\bar c,\bar d)$ commute if:
\begin{mathletters}
\begin{eqnarray}
\Delta(a,b,c,d) &=& \Delta(\bar a, \bar b, \bar c, \bar d) \\
\Gamma(a,b,c,d) &=& \Gamma(\bar a, \bar b, \bar c, \bar d) 
\end{eqnarray}
\end{mathletters}
with:
\begin{mathletters}
\label{e:gd}
\begin{eqnarray}
\Gamma(a,b,c,d) & = & {ab-cd \over ab + cd} \;,\\
\Delta(a,b,c,d) & = & {a^2+b^2-c^2-d^2 \over 2(ab+cd)} \;.
\end{eqnarray}
\end{mathletters}
Note that these necessary conditions
are valid for {\em any} lattice size $L$.
% vraiment????
One also gets the {\em same} conditions for the column-to-column
transfer matrices of this model.
Thus the commutation relations lead to
a foliation of the parameter space
in elliptic curves given by the intersection of
two quadrics Eq.~(\ref{e:gd}), that is
to an elliptic parameterization
(in the so-called principal regime \cite{Bax82}):
\begin{mathletters}
\label{e:parabax}
\begin{eqnarray}
a &=& \rho {\:{\rm sn}\,}(\eta-\nu)  \\
b &=& \rho {\:{\rm sn}\,}(\eta+\nu)  \\
c &=& \rho {\:{\rm sn}\,}(2\eta)  \\
d &=& -\rho\, k {\:{\rm sn}\,}(2\eta){\:{\rm sn}\,}(\eta-\nu)
{\:{\rm sn}\,}(\eta+\nu)
\end{eqnarray}
\end{mathletters}
where ${\:{\rm sn}\,}$ denotes the Jacobian elliptic function 
and $k$ their modulus.

It is also well known that the transfer matrix $T(a,b,c,d)$
commutes with the Hamiltonian of the anisotropic Heisenberg chain
\cite{Sut70}:
\begin{equation}
{\cal H} = -\sum_i J_x \sigma^x_i\sigma^x_{i+1}
 +  J_y \sigma^y_i\sigma^y_{i+1}
 +  J_z \sigma^z_i\sigma^z_{i+1}
\end{equation}
if:
\begin{equation}
\label{e:heisass}
1:\Gamma(a,b,c,d):\Delta(a,b,c,d) = J_x:J_y:J_z \;.
\end{equation}
This means that, given the three coupling constants $J_x$, $J_y$, and $J_z$
of a Heisenberg Hamiltonian, there exist infinitly many quadruplets
$(a,b,c,d)$ of parameters such that:
\begin{equation}
[T(a,b,c,d),{\cal H}(J_x,J_y,J_z)] = 0 \;.
\end{equation}
Indeed the three constants $J_x$, $J_y$, and $J_z$ determine uniquely
$\eta$ and $k$ in the elliptic parameterization (\ref{e:parabax}) and
the spectral parameter
$\nu$ can take any value, thus defining a continuous one-parameter family.
Not only $T$ and ${\cal H}$ commute for arbitrary values
of the parameter $\nu$, but ${\cal H}$ is also related to the
logarithmic derivative of $T$ at $\nu=\eta$.
In this work, we examine only regions
with the extra condition $c=d$ to ensure that $T$ is symmetric,
and thus that the spectrum is symmetric.
Using the symmetries of the eight-vertex model, one finds that
the  model $(a,b,c,d)$, with $c=d$  mapped into its principal
regime, gives a model $(\bar a, \bar b, \bar c,\bar d)$ with
$\bar a = \bar b$. In terms of the elliptic parameterization this means
${\:{\rm sn}\,}(\eta-\nu)={\:{\rm sn}\,}(\eta+\nu)$ or $\nu=0$.

In summary, in the continuous one-parameter family of
commuting transfer matrices $T(\nu)$
corresponding to a given value of $\Delta$ and $\Gamma$,
there are two special values of the spectral parameter $\nu$:
$\nu=\eta$ is related to the Heisenberg Hamiltonian
${\cal H}(1,\Gamma,\Delta)$, and for $\nu=0$
the transfer matrix $T(\nu)$ is symmetric
(up to a gauge transformation).

\subsubsection{Six-Vertex Model}

The six-vertex model 
is a special case of the eight-vertex model: one disallows the two
last vertex configurations of Fig.~\ref{f:vertices}, this means
$d=d'=0$. Both, the symmetric
and asymmetric  six-vertex models,
have been analyzed using the Bethe ansatz 
or also the Yang-Baxter equations
\cite{Bax82,LiWu72,Nol92}.
We did not examine this situation any further since
condition $c=d$ 
to have a real spectrum (see paragraph \ref{s:c=d}(i))
leads to a trivial case.

\subsubsection{Free-Fermion Condition}

Another case where the asymmetric eight-vertex model can be solved is
the case where the Boltzmann weights verify the 
so-called {\em free-fermion} condition:
\begin{equation}
\label{e:ff}
aa'+ bb' = cc'+ dd'
\end{equation}
For condition (\ref{e:ff}) the model reduces to
a quantum problem of free fermions
and the partition function can thus be computed 
\cite{FaWu70,Fel73c}.

The free-fermion asymmetric eight-vertex model 
is Yang--Baxter integrable, however the parameterization of the
Yang--Baxter equations is more involved
compared to the situation described in section \ref{ss:s8v}:
the row-to-row and column-to-column commutations
correspond to two different foliations of the parameter
space in algebraic surfaces.

It is also known that the asymmetric
eight-vertex free-fermion model can be mapped
onto a checkerboard Ising model. In
Appendix \ref{a:ff}  we give the correspondence between the vertex
model and the spin model. The partition function
per site of the model can be expressed in term of elliptic
functions $E$ which are not (due to the complexity
of the parameterization of the Yang--Baxter equations)
straightforwardly related to the two sets of surfaces
parameterizing the Yang--Baxter equations or even
to the canonical elliptic parameterization
of the generic (non free-fermion) asymmetric eight-vertex
model (see Eqs~\ref{e:paraas} in the following,
see also \cite{BeMaVi92}). The elliptic modulus
of these elliptic functions $E$ is given in Appendix \ref{a:ff}
as a function of the checkerboard Ising variables as well as
in the homogeneous Boltzmann weights 
($a$, $a'$, $b$, $b'$, $c$, $c'$, $d$, and $d'$) for the
free-fermion asymmetric eight-vertex model.

Finally, we remark that the restriction $c=d$ is compatible
with the condition (\ref{e:ff}) and, in contrast with the 
asymmetric six-vertex model, the asymmetric free-fermion model
provides a case where the row-to-row transfer matrix of the model
is symmetric.

\subsubsection{Disorder Solutions}

If the parameters $a$, $a^\prime$, $b$, $b^\prime$, $c$, and
$d$ are chosen such that the $R$-matrix (\ref{e:Rmat})
has an eigenvector which is a pure tensorial product:
\begin{equation}
\label{e:condeso}
\cal{R}  
\left( \begin{array}{c} 1 \\ p \end{array} \right)
\otimes
\left( \begin{array}{c} 1 \\ q \end{array} \right)
=
\lambda
\left( \begin{array}{c} 1 \\ p \end{array} \right)
\otimes
\left( \begin{array}{c} 1 \\ q \end{array} \right)
\end{equation}
then the vector:
\begin{equation} \label{e:vectorpq}
\left( \begin{array}{c} 1 \\ p \end{array} \right)
\otimes
\left( \begin{array}{c} 1 \\ q \end{array} \right)
\otimes
\cdots
\otimes
\left( \begin{array}{c} 1 \\ p \end{array} \right)
\otimes
\left( \begin{array}{c} 1 \\ q \end{array} \right)
\end{equation}
 ($2 L$ factors)
is an eigenvector of the diagonal(-to-diagonal) 
transfer matrix $\tilde T_L$, usually simply called the diagonal
transfer matrix.
The corresponding eigenvalue is $\Lambda=\lambda^{2L}$,
with
\begin{equation}
\lambda = {aa'-bb'+cc'-dd' \over (a+a')-(b+b')}  \; .
%= \frac{1}{2} \left( a+a'  + \sqrt{(a-a')^2+4 d^2}  \right)
\label{e:lambda}
\end{equation}
However, the eigenvalue $\Lambda$ may, or may not, be the eigenvalue
of largest modulus.
This corresponds to the existence of so-called
disorder solutions \cite{JaMa85} for which some dimensional
reduction of the model occurs \cite{GeHaLeMa87}.
Condition (\ref{e:condeso}) is simple
to express, it reads:
\begin{eqnarray}
\label{e:condeso2}
\lefteqn{A^2 + B^2 + C^2 + D^2 + 2 A B  - 2 A D - 2 B C - 2 C D= } \nonumber \\
& & (A+B-C-D) (a + b) (a^\prime + b^\prime)
-(A-D)(b^2 + b^{\prime 2}) - (B-C) (a^2 +a^{\prime 2})
\end{eqnarray}
where $A=aa^\prime$, $B=bb^\prime$, $C=cc^\prime$, and $D=dd^\prime$.
Note that in the symmetric case
 $a=a^\prime$, $b=b^\prime$, $c=c^\prime$, and
$d=d^\prime$, Eq. (\ref{e:condeso2}) factorizes as:
\begin{equation}
 (a - b + d - c) (a - b + d + c) (a - b - d - c) (a - b - d + c) = 0
\nonumber \end{equation}
which is the product of terms giving two disorder varieties
and two critical varieties of the Baxter model.
It is known that the symmetric model has four disorder varieties
(one of them, $a+b+c+d=0$, is not in the physical domain of
the parameter space)
 and four critical varieties \cite{Bax82}.
The missing varieties can be obtained by replacing ${\cal R}$ by
${\cal R}^2$ in Eq.~(\ref{e:condeso}).
In our numerical calculations we have always found 
for the asymmetric eight-vertex model that
 $\Lambda$ is either the eigenvalue of largest modulus
or  the eigenvalue of lowest modulus.
Finally, note that condition (\ref{e:condeso2}) does
{\em not} correspond to a solution of the Yang-Baxter equations.
This can be understood since 
disorder conditions like (\ref{e:condeso2}) are not
invariant under the action of the 
infinite discrete symmetry group $\Gamma$ presented in the next
subsection, whereas the solutions of the Yang-Baxter equations 
are actually invariant
under the action of this group
\cite{BMV:vert,Ma86}. 

On the other hand, similarly to the Yang--Baxter equations,
the ``disorder solutions''
can be seen to correspond to families of commuting
diagonal transfer matrices $\tilde T_L$ on a subspace $V$
of the $2^{2L}$ dimensional space on which $\tilde T_L$ acts:
\begin{equation}
[\tilde T_L(a,a',b,b',c,d),
  \tilde T_L(\bar a,\bar a',\bar b,\bar b',\bar c,\bar d)] \bigm|_V =0\;,
\end{equation}
where subscript $V$ means that the commutation is only valid on
the subspace $V$.
Actually, this subspace is the one-dimensional subspace corresponding
to vector (\ref{e:vectorpq}).
The notion of transfer matrices commuting only on a subspace $V$ can
clearly have precious consequences on the calculation of the
eigenvectors and eigenvalues, and hopefully of the partition function
per site.
One sees that the Yang--Baxter integrability and the disorder
solution ``calculability'' are two limiting cases where $V$
respectively corresponds to the entire space where $\tilde T_L$ acts
and to a single vector, namely Eq.~(\ref{e:vectorpq}).

\subsection{Some Exact Results on the Asymmetric Eight-Vertex Model}
\label{ss:exares}

When the Boltzmann weights of the model do not
verify any of the conditions of the preceding section, the 
partition function of the model has not yet been calculated.
However, some analytical results can be stated. 
Algebraic varieties of the parameter space can be presented,
which have very special symmetry properties.
The intersection of these algebraic varieties with critical
manifolds of the model are candidates for multicritical points
\cite{hm1}.
 We have tested the properties of the spectrum 
of the transfer matrices on these loci of the phase space.

There actually exists an infinite discrete group of symmetries
of the parameter space of the asymmetric eight-vertex model
(and beyond of the sixteen-vertex model \cite{BeMaVi92}). %aussi prl2
The critical manifolds of the model have to be compatible
with this group of symmetries and this is also true
for any exact property of the model: for instance
if the model is Yang--Baxter integrable, the YBE are compatible
with this infinite symmetry group 
\cite{BMV:vert}. %infinite ... vertex model.
However, it is crucial to recall that this symmetry group is not
restricted to the Yang--Baxter integrability. It is a symmetry group
of the model {\em beyond} the integrable framework and
provides for instance a canonical elliptic foliation of
the parameter space of the model
(see the concept of quasi-integrability \cite{BeMaVi92}).
The group is generated by simple transformations of the
homogeneous parameters of the model: the matrix inversion and
some representative geometrical symmetries, as
for example the geometrical symmetry of the
square lattice which amounts to
a simple exchange of $c$ and $d$:
\[
t_1 
 \left (
\begin {array}{cccc} 
a &0 &0 &d \\
0 &b &c &0 \\
0 &c &b'&0 \\
d &0 &0 &a'
\end {array}
\right )
=
 \left (
\begin {array}{cccc} 
a&0&0&c\\
0&b&d&0\\
0&d&b^\prime&0\\
c&0&0& a^\prime
\end {array}
\right )
\]
Combining  $I$ and $t_1$ yields an infinite discrete 
group $\Gamma$ of symmetries of the parameter space
\cite{Ma86}. 
This group is isomorphic to the infinite dihedral group
(up to a semi-direct product with ${\cal Z}_2$).
An infinite order generator of the non-trivial part of this group
is for instance $t_1 \cdot I $.
In the parameter space of the model this
generator yields an infinite set of points located on  elliptic curves.
The analysis of the orbits of the group $\Gamma$ for the 
asymmetric eight-vertex model yields (a finite set of) elliptic curves 
given by:
\begin{equation}
\label{e:paraas}
\frac{(a a'  +  b b'  -  c  c' - d  d')^2}{a a' b b'} ={\rm const}
,\quad \frac{a a' b  b'}{c c' d d'} = {\rm const}
\end{equation}
and
\[
\frac{a}{a'} = {\rm const} ,\quad
\frac{b}{b'} = {\rm const} ,\quad
\frac{c}{c'} = {\rm const} ,\quad
\frac{d}{d'} = {\rm const}.
\]
In the limit of the symmetric eight-vertex model one recovers the
well-known elliptic curves (\ref{e:gd}) of the Baxter model
given by the intersection of two quadrics.
Recalling  parameterization (\ref{e:parabax}) one sees that $t_1\cdot I$,
the infinite order generator of $\Gamma$, is actually
represented as a shift by $\eta$ of the spectral parameter:
$\nu \rightarrow \nu+\eta$.

The group $\Gamma$ is generically infinite, however, if some
conditions on the parameters hold, it degenerates into a finite
group. 
These conditions define algebraic varieties for which the model
has a high degree of symmetry. 
The location of multicritical points seems to correspond to
enhanced symmetries namely to the algebraic varieties where the
symmetry group $\Gamma$ degenerates into a finite group
\cite{hm1}. % hm six states Potts
Such conditions of commensuration of the shift $\eta$ with one
of the two periods of the elliptic functions occurred many times
in the literature of theoretical physics
(Tutte--Behara numbers, rational values of the central charge and
of critical exponents \cite{MaRa83}).
Furthermore, one can have, from the conformal field theory literature,
a prejudice of free-fermion parastatistics on these algebraic
varieties of enhanced symmetry \cite{DaDeKlaMcCoMe93}.
It is thus natural to concentrate on them.
We therefore have determined 
an {\em infinite} number of these algebraic varieties, which are remarkably
{\em codimension-one}  varieties of the parameter space.
Their explicit expressions
become quickly very large in terms of the homogeneous parameters
of the asymmetric eight-vertex model, however,
their expressions are remarkably simple in terms of 
some algebraic invariants generalizing those
of the Baxter model, namely:
\begin{mathletters}
\begin{eqnarray}
J_x & = & \sqrt{ a a'  b b'} + \sqrt{c c' d d'}  \\
J_y & = & \sqrt{ a a'  b b' }- \sqrt{c c' d d'}  \\
J_z & = &{{ a a' + b b' - c c' - d d'}\over {2}}.
\end{eqnarray}
\end{mathletters}
Note that, in the symmetric subcase, one recovers Eqs.~(\ref{e:heisass}).
In terms of these well-suited homogeneous variables,
it is possible to extend 
the ``shift doubling'' ($ \eta \rightarrow 2 \eta$) and 
``shift tripling'' ($ \eta \rightarrow 3 \eta$) transformations
of the Baxter model to the asymmetric eight-vertex model. 
One gets for the shift doubling transformation:
\begin{mathletters}
\label{doubling}
\begin{eqnarray}
J_x' & = & J_z^2 J_y^2 - J_x^2 J_y^2- J_z^2 J_x^2  \\
J_y' & = & J_z^2 J_x^2 - J_x^2 J_y^2- J_z^2 J_y^2  \\
J_z' & = & J_x^2 J_y^2 - J_z^2 J_x^2- J_z^2 J_y^2
\end{eqnarray}
\end{mathletters}
and for the shift tripling transformation:
\begin{mathletters}
\label{three}
\begin{eqnarray}
J_x''  & = & 
      \left (-2 J_z^2J_y^2J_x^4 -3 J_y^4J_z^4+2 J_y^2J_z^4J_x^2+J_y^4
	J_x^4+2 J_y^4J_z^2J_x^2+J_z^4J_x^4
\right ) \cdot  J_x
 \\
J_y''  & = & 
\left (2 J_z^2J_y^2J_x^4  
	-3 J_z^4J_x^4+J_y^4J_x^4 -2 J_y^4J_z^2J_x^2+J_y^4J_z^4  
	+2 J_y^2J_z^4J_x^2
\right ) \cdot J_y \\
J_z''  & = & 
\left (J_y^4J_z^4+2 J_y^4J_z^2J_x^2 -3 J_y^4J_x^4  
	-2 J_y^2J_z^4J_x^2+2 J_z^2J_y^2J_x^4  
	+J_z^4J_x^4
\right ) \cdot J_z
\end{eqnarray}
\end{mathletters}
The simplest codimension-one finite order varieties are:
$J_x=0$,  $J_y=0$, or $J_z=0$.
One remarks that $J_z=0$ is nothing but the free-fermion condition
(\ref{e:ff}) which is thus a condition for $\Gamma$ to be finite.
Another simple example is:
\begin{equation}
 J_y  J_z - J_x J_y - J_x J_z = 0,
\end{equation}
and the relations obtained by all permutations of $x$, $y$, and $z$.
Using the two polynomial transformations (\ref{doubling})
and  (\ref{three}) one can easily get an {\em infinite number}
of codimension-one algebraic varieties of finite order.
The demonstration that the codimension-one algebraic varieties
built in such a way are actually finite order conditions
of $\Gamma$ will be given elsewhere.
% It uses a compatibility
%of $\Gamma$ with an $sl_2\times sl_2\times sl_2\times sl_2$ group.
Some low order varieties are given in Appendix \ref{a:invariant}.
In the next section, the lower order varieties are tested
from the view point of statistical properties of the
transfer matrix spectrum.

%%%%%%%%%%%%%%%%%%%%%%%%%%%%%%%%%%%%%%%%%%%%%%%%%%%%%%%%%%%%%%%%%%%%%%%%
\section{Results of the RMT Analysis}
\label{s:results8v}

\subsection{General Remarks}

The phase space of the asymmetric eight-vertex model with
the constraint $c=d$ (ensuring symmetric transfer matrices
and thus real spectra)
is a four-dimensional space
(five homogeneous parameters $a$, $a'$, $b$, $b'$, and $c$).
Many particular algebraic varieties of this four-dimensional space
have been presented in the previous section and will now be
analyzed from the random matrix theory point of view.
We will present the full distribution of eigenvalue spacings
and the spectral rigidity at some representative points.
Then we will analyze the behavior of the eigenvalue spacing
distribution along different paths in the four-dimensional
parameter space.
These paths will be defined keeping some Boltzmann weights
constant and parameterizing the others by a single parameter $t$.

We have generated  transfer matrices
for various linear sizes, up to $L=16$ vertices,
leading to transfer matrices of size up to $65536\times65536$.
Tables~\ref{t:aj14} and \ref{t:aj16} give the dimensions of the different
invariant subspaces for $L=14$ and $L=16$.
Note that the size of the blocks to diagonalize increases
exponentially with the linear size $L$.
The behavior in the various subspaces is not significantly different.
Nevertheless the statistics is better for larger blocks since
the influence of the boundary of the spectrum 
and finite size effects are smaller. To get better statistics
we also have averaged the results of several blocks
for the same linear size $L$.

\subsection{Near the Symmetric Eight-Vertex Model}

\begin{figure}
\psfig{file=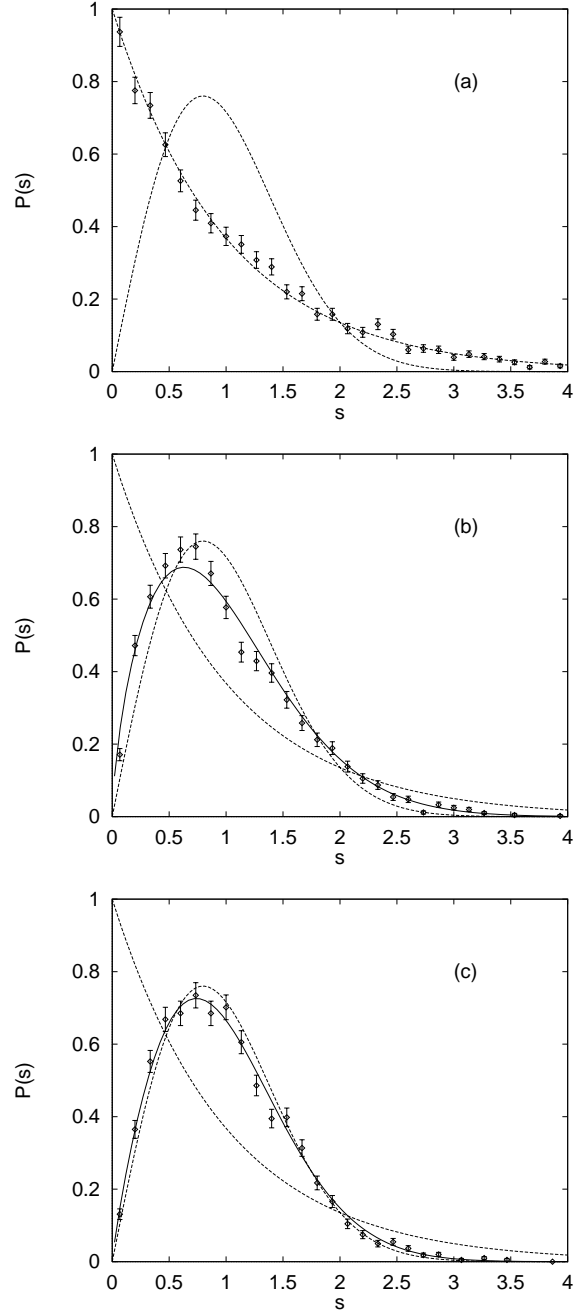,width=\hsize}
\caption{The distribution $P(s)$ of eigenvalue spacings for the three sets of
Boltzmann weights enumerated in Tab.~\protect\ref{t:pdscases}.
The continuous lines give the exponential for a Poissonian spectrum
and the Wigner distribution (\protect\ref{e:wigner}) for the GOE,
the broken lines in figures (b) and (c) are the fitted Brody distribution
with the $\beta$ given in Tab.~\protect\ref{t:pdscases}.}
\label{f:pds}
\end{figure}

Fig.~\ref{f:pds} presents the
probability distribution of the eigenvalue spacings for three
different sets of Boltzmann weights which are listed in
Tab.~\ref{t:pdscases}.
Fig.~\ref{f:pds}a) corresponds to a point of a symmetric eight-vertex model
while the other cases (b) and (c) are results for the asymmetric
eight-vertex model.
The data points result from about 4400 eigenvalue spacings coming from the
ten even subspaces for $L=14$ which are listed in Tab.~\ref{t:aj14}.
For the symmetric model (a),
using the symmetry under reversal of all arrows,
these blocks can once more be splitted into two sub-blocks of equal size.
The broken lines show the exponential and the Wigner distribution
as the exact results for the diagonal random matrix ensemble
(i.e.\ independent eigenvalues)
and the $2\times2$ GOE matrices.
In Fig.~\ref{f:pds}a) the data points fit very well an exponential,
whereas in Figs.~\ref{f:pds}b) and \ref{f:pds}c) 
they are close to the Wigner surmise.
In the latter cases we have also added the best fitting Brody distribution
with the parameter $\beta$ listed in Tab.~\ref{t:pdscases}.
The agreement with the Wigner distribution is better for the
case (c) where the asymmetry expressed by the ratio $a/a'$ is bigger.

\begin{figure}
\psfig{file=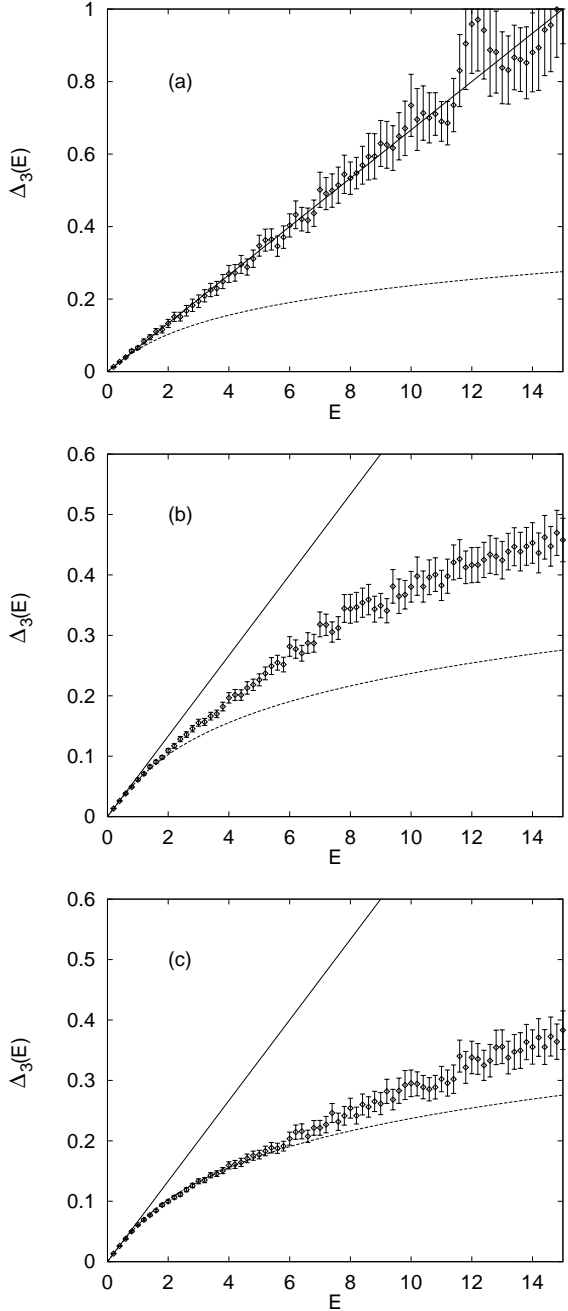,width=\hsize}
\caption{The spectral rigidity $\Delta_3(E)$ for the three sets of
Boltzmann weights enumerated in Tab.~\protect\ref{t:pdscases}.
The lines give the limiting case for a Poissonian spectrum and for
the GOE.}
\label{f:d3}
\end{figure}

We also have calculated the spectral rigidity to test how accurate
is the description of spectra of transfer matrices in terms of
spectra of mathematical random matrix ensembles.
We present in Fig.~\ref{f:d3}
the spectral rigidity $\Delta_3(E)$
for the same points in parameter space corresponding to integrability
and to non-integrability as in Fig.~\ref{f:pds}. The two
limiting cases corresponding to the Poissonian distributed
eigenvalues (solid line) and to GOE distributed eigenvalues (dashed line)
are also shown.
For the integrable point the agreement between the numerical
data and the expected rigidity is very good.
For the non-integrable case
the departure of the rigidity from the expected behavior
appears at $E \approx 2$ in case (b) and at $E\approx6$ in case
(c) (in units of eigenvalue spacings),
indicating that the RMT analysis is only valid at short scales.
Such behavior has already been seen in quantum spin systems
\cite{MoPoBeSi93,BrAdA96}.
We stress that the numerical results concerning the rigidity
depend much more on the unfolding than the results concerning
the spacing distribution.

Summarizing the results for the eigenvalue spacing distribution
and the rigidity, we have found very good agreement with the
Poissonian ensemble for the symmetric eight-vertex model (a),
good agreement with the GOE for the asymmetric model (c) and 
some intermediate behavior for the asymmetric eight-vertex model (b).
The difference between
the behavior for the cases (b) and (c) can be explained by
the larger asymmetry in case (c): case (b) is closer to
the integrable symmetric eight-vertex model.

%\subsection{Finite Size Behavior}

\begin{figure}
\psfig{file=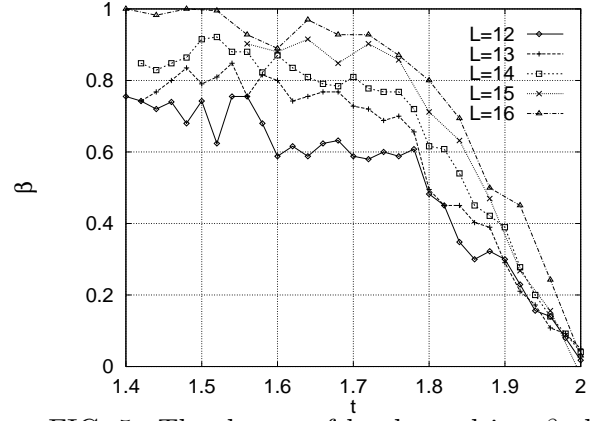,width=\hsize}
\caption{The degree of level repulsion $\beta$ obtained by a fit of the
eigenvalue spacing distribution $P(s)$ to the Brody distribution for
lattice sizes $L=12,\dots, 16$. The Boltzmann weights vary along the path
($a=t$, $a'=4/t$, $b=b'=4/5$, $c=d=\protect\sqrt{5/8}$) which
gives a symmetric model for $t=2$.
The number of spacings used for each size is given
in Tab.~\protect\ref{t:fss}.}
\label{f:fss}
\end{figure}

To study the proximity to the integrable model, we
have determined the `degree' of level repulsion $\beta$ by fitting the
Brody distribution to the statistics along a path
($a=t$, $a'=4/t$, $b=b'=4/5$, $c=\sqrt{5/8}$)
joining the cases (a) and (c) for different lattice sizes.
The result is shown in Fig.~\ref{f:fss}, the details
about the number of blocks and eigenvalue spacings used in the
distributions are listed in Tab.~\ref{t:fss}.
A finite size effect is  seen: we always find $\beta\approx0$ for
the symmetric model at $a/a'=1$ and increasing the system size
leads to a better coincidence with the Wigner distribution ($\beta=1$) for
the non-integrable asymmetric model $a\not=a'$.
So in the limit $L\rightarrow \infty$
we claim to find a delta-peak at the symmetric point $a=a'=2$.
We also have found that the size effects are really
controlled by the length $L$ and not by the size of the block.
However, our finite size analysis is only qualitative.
There is an uncertainty on $\beta$ of about $\pm0.1$.
There are two possible sources for this uncertainty.
The first one is a simple statistical effect and
could be reduced increasing the number of spacings.
The second one is a more inherent problem due to the
empirical parameters in the unfolding procedure. This
source of errors can not be suppressed increasing the size $L$.
For a quantitative analysis of the finite size effects
it would be necessary to have a high accuracy on $\beta$ and
to vary $L$ over a large scale, which is not possible because
of the exponential growth of the block sizes with~$L$.

To test a possible extension of the critical variety $a=b+c+d$
outside the symmetric region $a=a'$, $b=b'$,
we have performed similar calculations along the path
($a=t$, $a'=4/t$, $b=b'=4/5$, $c=3/5$) crossing the symmetric
integrable variety at a critical point $t=2$. The results are the same:
we did not find any kind of Poissonian behavior when $t \neq 2$.
%
%The critical manifold which extends the critical variety of
%the Baxter model is probably a transcendental manifold
%and not an algebraic variety (see Lieb and Wu in \cite{LiWu72}).
We have tested one single path and not the whole neighborhood
around the Baxter critical variety.
This would be necessary if one really wants to test a
possible relation between Poissonian behavior and criticity
instead of integrability
(both properties being often related).
The possible relation between Poissonian behavior
and criticity will be discussed for spin models in the second paper.

We conclude, from all these calculations, that the analysis
of the properties of the unfolded spectrum of
transfer matrices provides an efficient way to detect
integrable models, as already known for the
energy spectrum of quantum systems 
\cite{PoZiBeMiMo93,HsAdA93,MoPoBeSi93,vEGa94}.

\subsection{Case of Poissonian Behavior 
for the Asymmetric Eight-Vertex Model}

\begin{figure}
\psfig{file=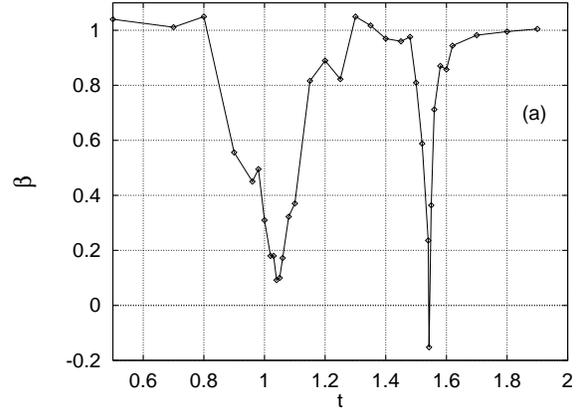,width=\hsize}
\caption{The degree of level repulsion $\beta$ for path (a) (see Tab.~\protect\ref{t:paths})
($L=16$ and 1000 spacings)}
\label{f:beta1}
\end{figure}

\begin{figure}
\psfig{file=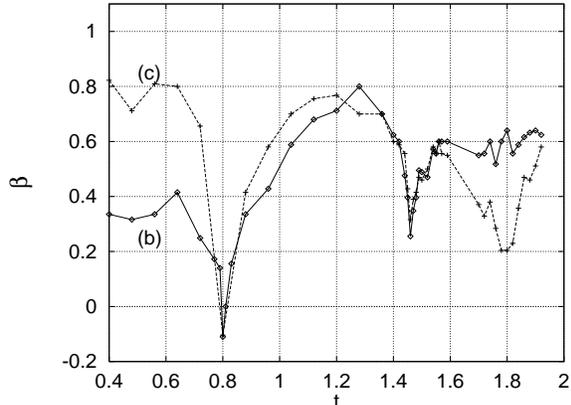,width=\hsize}
\caption{The degree of level repulsion $\beta$ for paths (b) and (c) (see Tab.~\protect\ref{t:paths})
($L=16$ and 1000 spacings for (b) and 3000 for (c))}
\label{f:beta3}
\end{figure}

\begin{figure}
\psfig{file=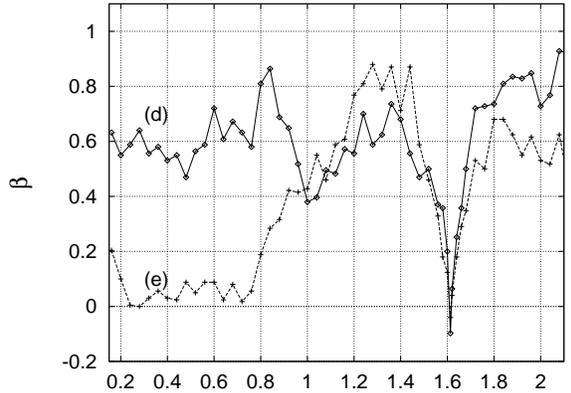,width=\hsize}
\caption{The degree of level repulsion $\beta$ for paths (d) and (e) (see Tab.~\protect\ref{t:paths})
($L=14$ and about 2000 spacings)}
\label{f:beta2}
\end{figure}

We now  investigate the phase space far from the Baxter model.
We define paths in the asymmetric region which cross the varieties
introduced above but which do not cross the Baxter model.
These paths and their intersection with the different varieties
are summarized in Tab.~\ref{t:paths}.
Fig.~\ref{f:beta1} corresponds to the path (a)
($a=4/5$, $a'=5/4$, $b=b'=t$, $c=1.3$).
This defines a path which crosses the free-fermion
variety at the solution of Eq.~(\ref{e:ff}): $t=t_{\rm ff}=\sqrt{2.38}$ 
and the disorder variety at
the two solutions of Eq.~(\ref{e:condeso}): $t=t_{\rm di}^{\rm max}\approx1.044$ and 
at $t=t_{\rm di}^{\rm min}\approx1.0056$ 
(the subscript ``di'' stands for disorder).
See Tab.~\ref{t:paths} for the intersections with the other varieties.
We have numerically found that, at the point $t=t_{\rm di}^{\rm max}$, the
eigenvalue (\ref{e:lambda}) is the one of largest modulus,
whereas at $t=t_{\rm di}^{\rm min}$ it is  
the eigenvalue of smallest modulus
(this is the superscript min or max).
The results shown are obtained using
the representation $R=0$ for $L=16$ (see Tab.~\ref{t:aj16}).
After unfolding and discarding boundary states we are left with
a distribution of about 1100 eigenvalue spacings.
One clearly sees that, most of the time, 
$\beta$ is of the order of one, signaling that the spacing
distribution is very close to the Wigner distribution,
except for $t=t_{\rm ff}\approx1.54$ and for $t$ close to 
the disorder solutions, where $\beta$ is close to zero.
This is not surprising for $t=t_{\rm ff}$
since the model is Yang--Baxter integrable at this point. 
The value $\beta(t_{\rm ff})$ is slightly negative: this
is related to a `level attraction' already noted in \cite{hm4}.
The downward peak is very sharp
and a good approximation of a $\delta$ peak that
we expect for an infinite size.
At $t=t_{\rm di}^{\rm min}$ and $t = t_{\rm di}^{\rm max}$
the model is {\em not} Yang-Baxter integrable.
We cannot numerically resolve between these two points.
Therefore, we now study paths where these two 
disorder solutions are clearly distinct.
For Fig.~\ref{f:beta2} they are both below the free-fermion point,
while for Fig.~\ref{f:beta3} the free-fermion point is between
the two disorder solution points.

In each of the Figs.~\ref{f:beta3} and \ref{f:beta2}
are shown the results for two paths 
which differ only by an exchange of the two weights $a\leftrightarrow a'$.
In Fig.~\ref{f:beta3}
one clearly sees a peak to $\beta$ slightly negative
at the free-fermion point at $t=0.8$
and another one at one disorder solution point 
$t=t_{\rm di}^{\rm max}\approx1.46$
for both curves but no peak at the second disorder solution
point $t=t_{\rm di}^{\rm min}\approx0.55$. 
It is remarkable that only point
$t=t_{\rm di}^{\rm max}$ yields the eigenvalue of largest modulus
for the diagonal(-to-diagonal) transfer matrix. Consequently,
one has the partition function per site of the model at this point.
At point $t=t_{\rm di}^{\rm min}$, 
where the partition function is not known,
we find level repulsion. However, only for path (c) the degree of
level repulsion $\beta$ is close to unity
while for path (b) it saturates at a much smaller value.
Another difference between the cases (b) and (c)
is a minimum in the curve
of $\beta(t)$ for path (c) at $t\approx 1.8$ which
is not seen for path (b). We do not
have a theoretical explanation for these phenomena yet:
these points are not located on any of the varieties presented
in this paper. We stress that an explanation cannot straightforwardly be 
found in the framework of the symmetry group $\Gamma$ presented here
since  $a$ and $a'$ appear only with the product $aa'$.
It also cannot be found in the Yang--Baxter framework,
since $a$ and $a'$ are on the same footing in the Yang--Baxter equations.

In Fig.~\ref{f:beta2} the curves of $\beta$ for the two paths (d) and (e)
again coincide very well at the free-fermion point
at $t=t_{\rm ff}\approx1.61$. 
But the behavior is very different for $t<1$ where the solutions
of the disorder variety are located.
For the path (d) neither of the two disorder points
is seen on the curve $\beta(t)$ which is almost 
stationary near a value around 0.6. This means that some
eigenvalue repulsion occurs, but the entire distribution is 
not very close to the Wigner surmise.
On the contrary, for path (e) the spacing distribution is very close to
a Poissonian distribution ($\beta(t) \approx 0$) 
when $t$ is between the two disorder points.
This suggests that the status of eigenvalue spectrum on
the  disorder variety of the asymmetric eight-vertex model is not simple:
a more systematic study would help to clarify the situation.

We now summarize the results from the Figs.~\ref{f:beta1}--\ref{f:beta2}:
generally, the statistical properties of the transfer matrix spectra
of the asymmetric eight-vertex model are close to those of the GOE
except for some algebraic varieties.
We have always a very sharp peak with $\beta\rightarrow 0$
at the free-fermion point, often $\beta\approx-0.2$.
All other points with $\beta \rightarrow 0$ are found
to be a solution of the  disorder variety (\ref{e:condeso2})
of the asymmetric eight-vertex model.

\subsection{Special Algebraic Varieties} 

To conclude this section we discuss the special
algebraic varieties of the symmetry group $\Gamma$. As explained
in subsection \ref{ss:exares} it is possible to construct
an infinite number of algebraic varieties where the generator
is of finite order $n$: $(t_1\cdot I)^n = {\rm Id}$
and thus $\Gamma$ is finite order.
As an example, the solutions for $n=6$ and $n=16$ are given in Appendix
\ref{a:invariant}. We have actually calculated a third variety,
the expression of which is too long to be given ($n=8$). 
We give in Tab.~\ref{t:paths} the values of the parameter
$t$ for which each path crosses each variety
$t_{\rm fo}^6$, $t_{\rm fo}^8$, $t_{\rm fo}^{16}$
(the subscript ``fo'' stands for finite order and the superscript
is the order $f$). It is easy to verify on the 
different curves that no tendency to  Poissonian behavior
occurs at these points.
We therefore give a negative answer to the question of a
special status of {\em generic} points of 
the algebraic finite order varieties
with respect to the statistical properties of
the transfer matrix spectra.
However, one can still imagine that subvarieties of these
finite order varieties could have Poissonian behavior and
be candidates for free parafermions or multicritical points.

\section{Conclusion and Discussion}

We have found that the entire spectrum of the symmetric
row-to-row transfer matrices of the eight-vertex model
of lattice statistical mechanics
is sensitive to the Yang--Baxter integrability of the model.
The GOE provides a satisfactory 
description of the spectrum of non Yang--Baxter
integrable models: the eigenvalue spacing distribution
and the spectral rigidity up to an energy scale
of several spacings are in agreement with the 
Wigner surmise and the rigidity of the
GOE matrix spectra. This accounts for
``eigenvalue repulsion''.
In contrast, for Yang--Baxter integrable
models, the unfolded spectrum has many features of a set of
independent numbers: the spacing distribution is Poissonian and the 
rigidity is linear over a large energy scale.
This accounts for ``eigenvalue independence''.
However, we have also given a non Yang--Baxter integrable
disorder solution of the asymmetric eight-vertex model.
For some parts of it the spectrum is clearly Poissonian, too.
This suggests that the Wignerian nature of the spectrum is not completely
controlled by the Yang--Baxter integrability alone, but possibly
by a more general notion of ``calculability'',
possibly based on the existence of a family of
transfer matrices commuting on the same subspace.
We have also found some ``eigenvalue attraction'' for some
Yang--Baxter integrable models, namely for most points of the
free-fermion variety. 
These results could surprise
since we do not a priori expect properties
involving all the $2^L$ eigenvalues when only the 
few eigenvalues of larger modulus have a physical significance. 
However, the eigenvalues of small modulus control the finite
size effects, and it is well known that, for example,
the critical behavior (critical exponents) can be deduced
from the finite size effects.
The nature of the eigenvalue spacing distribution being an effective
criterion, 
%This criterion of the nature of the eigenvalue spacing
%distribution being effective
we have also used it to
test unsuccessfully various special manifolds including
the vicinity of the critical variety of the Baxter model.
We will present in a forthcoming publication a similar
study of  spin model (rather than vertex models).
In particular it is interesting to  analyze 
the spectrum on a critical, but not Yang--Baxter integrable,
algebraic variety of codimension one
as it can be found in the $q=3$ Potts model on a triangular
lattice with three-spin interactions \cite{WuZi81}.
However, this leads models the transfer
matrix of which cannot be made symmetric.
This will require a particular study of complex spectra which is 
much more complicated.
In particular the eigenvalue repulsion becomes two-dimensional,
and to investigate the eigenvalue spacing distribution, one has
to analyze the distances between eigenvalues in two dimensions.

\acknowledgments
We would like to thank Henrik Bruus
for many discussions concerning random matrix theory.

\appendix
\section{}
\label{a:ff}
We give hereafter the correspondence between
the asymmetric eight-vertex model on the free-fermion variety
and the checkerboard Ising model.
The vertex model is specified by eight homogeneous parameters.
The gauge invariance (\ref{e:gauge}),
together with the free-fermion condition (\ref{e:ff}),
leads to only four independent parameters.
The checkerboard model is specified 
by the four independent usual low-temperature variables 
$x_i=e^{-K_i}$.

\begin{eqnarray}
\label{abcdcheck}
a   =   \frac{1}{2}  \left( x_ ,x_2 x_3 x_4\,
 + \frac{1}{ x_1 x_2 x_3 x_4}\right)\;,
 & & \quad
a'  =  \frac{1}{2}  \left(\frac{x_1 x_3}{ x_2 x_4}
+ \frac{x_2 x_4}{ x_1 x_3}\right)\\
 b  =  \frac{1}{2}  \left(\frac{x_1 x_4}{ x_2 x_3}
+ \frac{x_2 x_3}{ x_1 x_4} \right) \;,
 & & \quad
b'  =  \frac{1}{2}   \left(\frac{x_1 x_2}{ x_4 x_3}
+\frac{x_4 x_3}{ x_1 x_2}\right)  \\
c  =  \frac{1}{2}   \left(\frac{x_1 x_2 x_3}{ x_4}
+ \frac{x_4}{ x_1 x_2 x_3}\right)\;,
 & & \quad
c'  =  \frac{1}{2}   \left(\frac{x_1 x_4 x_3}{ x_2}
+ \frac{x_2}{ x_1 x_4 x_3}\right) \\
d  =  \frac{1}{2}   \left(\frac{x_4 x_3 x_2}{ x_1}
+ \frac{x_1}{ x_4 x_3 x_2}\right) \;,
 & & \quad
d'  =  \frac{1}{2}  \left(\frac{x_1 x_4 x_2}{ x_3}
+ \frac{x_3}{ x_1 x_4 x_2}\right)
\end{eqnarray}
The modulus $k_{\rm check}$ of the checkerboard
Ising model or equivalently of the asymmetric
eight-vertex model reads :
\begin{eqnarray}
k_{\rm check}^2 & = & {{k_N} \over {k_D}} 
\end{eqnarray}
with :
\begin{eqnarray*}
k_N & = & 16 \,( x_3^2 x_2^2 x_1^2 + x_4^2) 
\times \mbox{circular permutations}
\\
k_D & = &
( x_1^2 x_2^2 x_3^2 x_4^2 - x_1^2 x_2^2 - x_1^2 x_3^2
 + x_1^2 x_4^2 + x_3^2 x_4^2 + x_4^2  x_2^2 - x_3^2 x_2^2 - 1) \times
\mbox{(circular permutations)}
\end{eqnarray*}
or equivalently:
\begin{equation}
\label{modul}
k_{\rm free}^2  = 
\frac{16 \,c' d' c d}{16 c' d' c d-8 a' b a b'-2 a'^2 b^2
-2 a^2  b'^2 - 2 a^2  b^2 -
2 b'^2  a'^2 +
(a^2-a'^2)^2 + (b^2-b'^2)^2}
\end{equation}

\section{}
\label{a:invariant}

We give in this appendix some algebraic varieties
where the group $\Gamma$ becomes {\em finite}.
Two varieties of order six ($(t_1\cdot I)^6={\rm Id}$):
\begin{equation}
 a_\pm = \frac{c d^3\pm c^2 d^2+d c^3-c d b b'}{a' d c\pm a' b b'}
\end{equation}
A variety of order sixteen ($(t_1\cdot I)^{16}={\rm Id}$):
\begin{eqnarray}
(u^2-v^2)(u-v)+u v (u^2+v^2) & = & 0
\end{eqnarray}
with:
\begin{eqnarray}
u &=& \frac{c c' d d' (a a'+b b'-c^2-d^2)^2}{(a a' b b-c c' d d')^2} -1 
= \left( \frac{J_x J_z - J_y J_z}{J_x J_y} \right)^2 -1 \\
v &=& \frac{a a' b b' (a a'+b b'-c^2-d^2)^2}{(a a' b b-c c' d d')^2} -1 
= \left( \frac{J_x J_z + J_y J_z}{J_x J_y} \right)^2 -1 
\end{eqnarray}

%\bibliographystyle{/usr1/local/lib/tex/inputs/prsty}
%\bibliographystyle{prsty1}
%\bibliography{/home/DISK/HMEYER/BIBLIO/all,/home/DISK/HMEYER/BIBLIO/JM}

\begin{table}
\begin{tabular}{cccccc}
 R & $k$ & $\lambda$&  $l_R$ &  $a^{\rm even}_R$  & $a_R^{\rm odd}$ \\
\hline
0 & 0 & 1&1 &362 & 325 \\
1 & $\pi$ & $-1$&1 &288 & 325 \\
2 & 0 & $-1$&1 &234 & 261\\
3 & $\pi$ & 1&1 &288 & 261 \\
4 & $2\pi/7$ & *&2 &594 & 585\\
5 & $4\pi/7$ & *&2 &594 & 585 \\
6 & $6\pi/7$ & *&2 &594 & 585 \\
7 & $\pi/7$ & *&2 &576 & 585  \\
8 & $3\pi/7$ & *&2 &576 & 585  \\
9 & $5\pi/7$ & *&2 &576 & 585  
\end{tabular} %
\caption{The dimensions $a_R$ and degeneracies $l_R$ of the
invariant subspaces for $L=14$.
$R$ is an arbitrary label of the representations of
the dihedral group, $\exp(ik)$ and $\lambda$ are the eigenvalues
of the corresponding translation and reflection operators
(* means that the corresponding representation is not
stable under the action of the reflection operator).
The two numbers  $a_R^{\rm even}$ and $a_R^{\rm odd}$ correspond to the
blocks between configurations with an even or odd number of up arrows
(see text).}
\label{t:aj14}
\end{table}

\begin{table}
\begin{tabular}{cccccc}
 R & $k$ & $\lambda$ & $l_R$ &  $a^{\rm even}_R$  & $a_R^{\rm odd}$ \\
\hline
0 & 0     & 1 &1&1162 & 1088\\ % 0 & + \\
1 & $\pi$ & $-1$&1 &1033 & 1088\\ % $\pi$ & + \\
2 & 0     & $-1$&1 &  906 & 960\\ % 0 & $-$ \\
3 & $\pi$ &  1&1 &1033 & 960\\ % $\pi$ & $-$ \\
4 & $\pi/2$ & *&2 &2065 & 2048\\ % $\pi & 2$ & $\pm$ \\
5 & $\pi/4$ & *&2 &2062 & 2048\\ % $\pi/4$ & $\pm$ \\
6 & $3\pi/4$& *&2 &2062 & 2048\\ % $3\pi/4$ & $\pm$ \\
7 & $ \pi/8$ & *&2 &2032 & 2048\\ % $\pi/8$ & $\pm$ \\
8 & $7\pi/8$ & *&2 &2032 & 2048\\ % $3\pi/8$ & $\pm$ \\
9 & $5\pi/8$ & *&2 &2032 & 2048\\ % $5\pi/8$ & $\pm$ \\
10& $3\pi/8$ & *&2 &2032 & 2048 % $7\pi/8$ & $\pm$
\end{tabular} %
\caption{The parameters of the invariant subspaces as in the preceding
table for $L=16$.}
\label{t:aj16}
\end{table}

\begin{table}
\begin{tabular}{cccccc}
case & $a$ & $a'$ & $b=b'$ &$c=d$ & $\beta$ \\ \hline
a) & 2 & 2 & 0.8 & $\sqrt{5/8}\approx0.790569$ & 0.02 \\
b) & 1.8 & 2.2222 & 0.8 & $\sqrt{5/8}\approx0.790569$ & 0.64 \\
c) & 1.6 & 2.5 & 0.8 & $\sqrt{5/8}\approx0.790569$ & 0.85 
\end{tabular}
\caption{The Boltzmann weights for the three cases of figures
\protect\ref{f:pds} and \protect\ref{f:d3}.
The last column indicates
the parameter $\beta$ of the Brody distribution 
shown in Fig.~\protect\ref{f:pds}.}
\label{t:pdscases}
\end{table}

\begin{table}
\begin{tabular}{clc}
$L$ & dimension $a_j$ of the blocks& \#(spacings) \\ \hline
12 & 122, 176, 174, 166, 165,165 & 870 \\
13 & 190, 315, 315, 315 & 1000 \\
14 & 362, 596, 596, 576 & 2000 \\
15 & 612, 1092, 1094 & 2600 \\
16 & 1162, 1092, 2065 & 4000
\end{tabular}
\caption{Numbers of representations and their dimensions used for the
distributions leading to Fig.~\protect\ref{f:fss}.}
\label{t:fss}
\end{table}

\begin{table}
\begin{tabular}{lccccc}
 & (a) & (b) & (c) & (d) & (e) \\ \hline
$a$ & 0.8 & 0.4 & 2 & 1.2 & 0.8 \\ 
$a'$ & 1.25 &  2 & 0.4 & 0.8 & 1.2 \\ 
$b$ & $t$ & $t$ & $t$ & $t$ & $t$ \\ 
$b'$ & $t$ & 1.5 & 1.5 & 1.5 & 1.5\\ 
$c=d$ & 1.3 & 1 & 1 & 1.3 & 1.3 \\  \hline
$t_{\rm ff}$ & 1.5427 & 0.8 & 0.8 & 1.6133 & 1.6133 \\ 
$t_{\rm di}^{\rm min}$ & 1.0057 & 0.5520 & 0.5520 & 0.6156 & 0.6156 \\ 
$t_{\rm di}^{\rm max}$ & 1.0443 & 1.4608 & 1.4608 & 0.2424 & 0.2424 \\ 
$t_{\rm fo}^6$ & 1.3 & 0.6667& 0.6667 & 1.1267 & 1.1267 \\ 
$t_{\rm fo}^6$ & 1.5991 & 0.8141 & 0.8141 & 1.7474 & 1.7474 \\ 
$t_{\rm fo}^8$ & 1.6000 & 0.8141 & 0.8141 & 1.7001 & 1.7001 \\ 
$t_{\rm fo}^{16}$ &     & 0.5237 & 0.5237 & 0.6337 & 0.6337 \\ 
\end{tabular}
\caption{The parameterization of the paths (a)--(e) and their intersection
points with the different algebraic varieties (see text).}
\label{t:paths}
\end{table}

%%%%%%%%%%%%%%%%%%%%%%%%%%%%%%% FIGURES %%%%%%%%%%%%%%%%%%%%%%%%%%%%%%%

%% FOR SUBMISSION, COMMENT THE FOLLOWING LINE AND THE 7 lines "\psfig{}"
\twocolumn \input{psfig.tex}

\end{document}